\documentclass[12pt]{amsart}
\usepackage{amsmath,amsfonts,amssymb} \usepackage{color}
\usepackage[all,cmtip]{xy}
\usepackage{graphicx}
 \usepackage{youngtab}

\newcommand{\corr}[1]{\langle {#1} \rangle}

\newcommand{\ba}{{\bf a}}

  \newcommand{\cF}{\mathcal{F}} 
 
 \newcommand{\bT}{{\bf T}}

   \newcommand{\fg}{\mathfrak{g}}
  
 \newcommand{\bZ}{\mathbb{Z}}  \newcommand{\bA}{\mathbf{A}}
 \newcommand{\bC}{\mathbb{C}}
 \newcommand{\pd}{\partial}

\newcommand{\vac}{|0\rangle}  \newcommand{\lvac}{\langle 0|}

  \DeclareMathOperator{\res}{res}  
  
 \DeclareMathOperator{\gl}{gl}
\DeclareMathOperator{\Gr}{Gr}

\newcommand{\be}{\begin{equation}}
\newcommand{\ee}{\end{equation}}
\newcommand{\bea}{\begin{eqnarray}}
\newcommand{\eea}{\end{eqnarray}}
\newcommand{\ben}{\begin{eqnarray*}}
\newcommand{\een}{\end{eqnarray*}}

\newcommand{\half}{\frac{1}{2}}

\newtheorem{cor}{Corollary}[section]

 \newtheorem{thm}[cor]{Theorem}
\theoremstyle{remark}

\definecolor{A}{rgb}{.75,1,.75}

\definecolor{green}{rgb}{0,1,0}
\definecolor{yellow}{rgb}{1,1,0}
\definecolor{orange}{rgb}{1,.7,0}
\definecolor{red}{rgb}{1,0,0}
\definecolor{white}{rgb}{1,1,1}

 \makeindex
\begin{document}
\title 
{Fermionic Computations for Integrable Hierarchies}

\author{Jian Zhou}
\address{Department of Mathematical Sciences\\Tsinghua University\\Beijng, 100084, China}
\email{jzhou@math.tsinghua.edu.cn}

\begin{abstract}
We present a unified fermionic approach to compute the tau-functions and
the $n$-point functions of integrable hierarchies
related to some infinite-dimensional Lie algebras and their representations.
\end{abstract}

\maketitle

\section{Introduction}

In this paper we present a unified approach to compute the tau-functions and
the $n$-point functions of integrable hierarchies
related to some infinite-dimensional Lie algebras and their representations.
Our motivation is to generalize the method to obtain explicit fermionic expressions
for Witten-Kontsevich tau-function \cite{Zhou-Fermionic} 
and for the tau-function \cite{Balogh-Yang-Zhou} 
for intersection numbers on moduli spaces Witten's r-spin curves.

Our basic strategy is to use reductions of the KP hierarchy
and base on an earlier work \cite{Zhou-Emergent} on the KP hierarchy,
and a generalization of the method of Kac-Schwarz \cite{Kac-Schwarz}.
In \cite{Sato},
Sato introduced an infinite-dimensional Grassmannian (GM) as the moduli space
of solutions of the KP hierarchy.
The Sato Grassmannian was identified with the orbit space of the vacuum vector
in a fermionic Fock space under an action of the Kac-Moody group $GL(\infty)$.
By restricting to suitable subgroups of this group,
one obtains reductions of the KP hierarchy.
For example, the famous KdV hierarchy can be recovered in this way.
He conjectured that ``any soliton equations, or completely integrable systems, is obtained in this way."
He then proposed that ``Classification of soliton equations would then be reduced to classification of submanifolds
of our GM which are stable by the subgroup of $GL(\infty)$".
Denote by $A_\infty$ the Lie algebra of $GL(\infty)$.
This proposal was carried out by the Kyoto school \cite{DJKM, Jimbo-Miwa},
where the cases of  $B_\infty$, $C_\infty$, $D_\infty$ and the Kac-Moody algebras
$A_n^{(1)}$, $A_{n}^{(2)}$, $C_n^{(1)}$, $D_{n}^{(1)}$, $D_n^{(2)}$
were shown to be Lie subalgebras of $A_\infty$.
We will recall them in \S \ref{sec:DJKM}.
These are the cases where our method readily applies.
These subalgebras of $A_\infty$ are the fixed point sets of some automorphisms of $A_\infty$.
We will present some more constructions in \S \ref{sec:Matrix} 
based on affinization of embeddings of finite-dimensional Lie algebras.
The following are the salient features of the approach to integrable hierarchies of Japanese school:
tau-function, vertex operator construction of representations of
infinite-dimensional Lie algebras, and boson-fermion correspondence.

When restricted to these Kac-Moody algebras obtained by the Japanese school,
the Fock space representation induces representations of these Kac-Moody algebras \cite{KKLW},
and often these are the basic representations.
Generalizing to exceptional Lie algebras,
Kac and Wakimoto \cite{Kac-Wakimoto} constructed the hierarchies associated
to arbitrary loop groups in a uniform way.
Their construction associates an integrable hierarchy to
an affine Kac-Moody algebra $\fg$,
together with a vertex operator construction
$R$ of an integrable highest weight representation $V$ of $\fg$.
Therefore,
their construction further clarifies the role of explicit realizations of representations
in the theory of soliton systems.
Drinfeld and Sokolov \cite{Drinfeld-Sokolov}
gave a different construction based on the Zakharov-Shabat zero-curvature equation  
(see also \cite{deG-H-M} for generalizations).
This construction is more geometrical in the sense that the hierarchies are related to
bihamiltonian structures.
More recently there have appeared some constructions of integrable hierarchies
based on the theory of Frobenius manifolds by the work of Dubrovin-Zhang \cite{Dubrovin-Zhang}
and Givental-Milanov \cite{Givental-Milanov}.
They are inspired by Witten Conjecture/Kontsevich Theorem.
They have applications in FJRW theory \cite{FJR, LRZ}.
Many of these integrable hierarchies have been shown to be equivalent to each other,
and in particular to those obtained by reductions of the KP hierarchy,
hence our results can be applied to them.

We arrange the rest of the paper as follows.
In Section 2 we recall the general method for KP hierarchy developed in \cite{Zhou-Emergent}.
In Section 3 we recall the construction of the Japanese school of subalgebras of $A_\infty$.
In Section 4 we use affinization of embedding of Lie algebras to obtain more example.
In the final Section 5 are some concluding remarks.

\vspace{.3in}

{\em Acknoledgements}.
This research is partially supported by NSFC grant 11171174.  
Communications and collaborations with Professors Ference Balogh and Di Yang 
on a related problem \cite{Balogh-Yang-Zhou} are very helpful for this work.

\section{General Results on Tau-Function of KP Hierarchy} \label{sec:KP}

In this Section we will recall some results in \cite{Zhou-Emergent}
based on the work of Kyoto school on KP hierarchy.

\subsection{Sato's Grassmannian and semi-infinite wedge product} \label{sec:Sato}

Let $H$ be the space   consisting of  the  formal  Laurent  series
$\sum_{n \in \bZ} a_n z^{n-1/2}$,
such that $a_n =  0$  for  $n \gg 0$,
and let $H_+ = \{\sum_{n \geq 1} a_n z^{n-1/2} \in H\}$,
$H_- = \{ \sum_{n \leq 0} a_n z^{n-1/2} \in H\}$.
Then one has a decomposition:
\be
H = H_+ \oplus H_-.
\ee
Denote by $\pi_\pm: H \to H_\pm$ the natural projections onto these subspaces.
The big cell of Sato Grassmannian $\Gr_{(0)}$ consists of linear subspaces $U \subset H$ such that
$\pi_+|_U: U \to H_+$ is an isomorphism.

One can see that every $U \in \Gr_{(0)}$ has a  basis
of the form
\be
f_n = z^{n+1/2} + \sum_{m \geq 0} a_{n,m} z^{-m - 1/2},
\ee
called a normalized basis.
The coefficients $\{a_{n,m}\}$ are called the affine coordinates on the big cell \cite{Balogh-Yang}.
Given such a basis, one has
\ben
|U\rangle :& = & f_1 \wedge f_2 \wedge \cdots \\
& = & \sum \alpha_{m_1, \dots, m_l; n_1, \dots, n_l}\cdot z^{-m_1-1/2} \wedge \cdots \wedge z^{-m_l-1/2} \\
&& \wedge z^{1/2} \wedge \cdots \widehat{z^{n_l+1/2}} \wedge \cdots \wedge \widehat{z^{n_1+1/2}} \wedge \cdots,
\een
where $m_1> m_2 > \cdots > m_l \geq 0$, $n_1 > n_2 > \cdots > n_l \geq 0$ are two sequences of integers,
and
\be
\alpha_{m_1, \dots, m_l; n_1, \dots, n_l}
= (-1)^{n_1+ \cdots +n_l} \begin{vmatrix}
a_{n_1, m_1} & \cdots & a_{n_1, m_l} \\
\vdots & & \vdots \\
a_{n_l, m_1} & \cdots & a_{n_l, m_l}
\end{vmatrix}.
\ee

\subsection{Creators and annihilators on  fermionic Fock space  $\mathcal{F}$}

For a sequence $\ba = (a_1, a_2, \dots)$ of half-integers such that $a_1 < a_2 < \cdots$.
We say $\ba$ is admissible if both of  the sets $(\bZ_{\geq 0} + \half) - \{a_1, a_2, \dots\}$
and $\{a_1, a_2, \dots\} - (\bZ_{\geq 0} + \half)$
are finite.
For an admissible sequence $\ba$,
let
\be
|\ba\rangle : = z^{a_1} \wedge z^{a_2} \wedge \cdots .
\ee
The fermionic Fock space $\cF=\Lambda^{\frac{\infty}{2}}(H)$ is the space of expressions of form:
\be
\sum_{\ba} c_\ba |\ba\rangle,
\ee
where the sum is taken over admissible sequences.

As in the case of ordinary Grassmann algebra,
one can consider exterior products and inner products.
For $r \in \mathbb{Z}+\frac{1}{2}$,
define operator $\psi_r: \Lambda^{\frac{\infty}{2}}(H) \to \Lambda^{\frac{\infty}{2}}(H)$ by
\be
\psi_r |\ba\rangle = z^{r} \wedge |\ba\rangle,
\ee
and let $\psi_r^*: \Lambda^{\frac{\infty}{2}}(H) \to \Lambda^{\frac{\infty}{2}}(H)$ be defined by:
\be
\psi^*_r |\ba\rangle =
\begin{cases}
(-1)^{k+1} \cdot z^{a_1}\wedge\cdots\wedge \widehat{z^{a_k}}\wedge\cdots, & \text{if $a_k = -r$ for some $k$}, \\
0, &  \text{otherwise}.
\end{cases}
\ee
The anti-commutation relations for these operators are
\begin{equation} \label{eqn:CR}
\psi_r\psi^*_s + \psi^*_s\psi_r = \delta_{-r,s}id
\end{equation}
and other anti-commutation relations are zero.

The fermionic vacuum vector is
\be
\vac := z^{1/2} \wedge z^{3/2} \wedge \cdots.
\ee
It is clear that for $r > 0$,
\begin{align}
\psi_{r} \vac & = 0, & \psi_r^* \vac & = 0.
\end{align}
The operators $\{\psi_{r}, \psi_r^*\}_{r > 0}$ are called the fermionic annihilators,
and the operators $\{\psi_{r}, \psi_r^*\}_{r < 0}$ are called the fermionic creators.

\subsection{Tau-function of KP hierarchy}

The result in \S \ref{sec:Sato} can be reformulated as follows:

\begin{thm} \label{thm:Bogoliubov} (\cite[Therem 3.1]{Zhou-Emergent})
Suppose that $U$ is given by a normalized basis
$$\{f_n = z^{n+1/2} + \sum_{m \geq 0} a_{n,m} z^{-m - 1/2} \},$$
then one has
\be
|U\rangle = e^A \vac,
\ee
where $A: \cF \to \cF$ is a linear operator
\be
A = \sum_{m, n \geq 0} a_{n,m} \psi_{-m-1/2} \psi^*_{-n-1/2}.
\ee
\end{thm}

For a linear operator $L: \cF \to \cF$,
one can define its vacuum expectation value:
\be
\corr{L} : = \lvac L \vac.
\ee
One also defines
\be
\corr{L}_U : = \lvac L |U\rangle.
\ee
Consider the fermionic fields
\bea
&& \psi(\xi) = \sum_{r\in \half + \bZ} \psi_r \xi^{-r-1/2}, \\
&& \psi^*(\xi) = \sum_{r\in \half + \bZ} \psi_r^* \eta^{-r-1/2}.
\eea
It is easy to see that
\be
\corr{\psi (\xi) \psi^*(\eta)}_U
= i_{\xi, \eta} \frac{1}{\xi-\eta} + A(\xi, \eta),
\ee
where
\be
i_{\xi, \eta} \frac{1}{\xi-\eta} = \sum_{n \geq 0} \xi^{-n-1} \eta^n,
\ee
and
\be
A(\xi, \eta) =  \sum_{m,n\geq 0} a_{m,n} \xi^{-m-1} \eta^{-n-1}.
\ee
In particular,
$\corr{\psi (\xi) \psi^*(\eta)}_U$ contains the same information as the operator $A$.
Also define a bosonic field $\alpha(\xi)$ by
\be
:\psi(\xi)\psi^*(\xi): = \sum_{n \geq 0} \alpha_n \xi^{-n-1}.
\ee
The operators $\{\alpha_n\}_{n \in \bZ}$ satisfy the Heisenberg commutation relations:
\be
[\alpha_m, \alpha_n] = m \cdot \delta_{m, -n}.
\ee
The operator $\alpha_0$ is called the charge operator.
Let
\be
\cF_{n} = \{ v\in \cF \;|\; \alpha_0(v) = n \cdot v\}.
\ee
Then one has a charge decomposition:
\be
\cF = \bigoplus_{n\in \bZ} \cF_{n}.
\ee
The Sato tau-function associated to $U$ is defined by:
\ben
\tau_U(\bT) =\lvac e^{\sum\limits_{n \geq 1} T_n \alpha_n}|U\rangle  .
\een
It is a tau-function of the KP hierarchy.
Also define the free energy $F_U$ by:
\be
F_U(\bT) = \log \tau_U(\bT).
\ee

\begin{thm} \label{thm:Bosonic-N-Point} (\cite[Therem 5.3]{Zhou-Emergent})
For $n \geq 2$,
\be
\begin{split}
& \sum_{j_1,\dots, j_n \geq 1}
\frac{\pd^n F_U}{\pd T_{j_1} \cdots \pd T_{j_n} } \biggl|_{\bT =0}
  \xi_1^{-j_1-1}\cdots \xi_n^{-j_n-1} \\
= & (-1)^{n-1} \sum_{\text{$n$-cycles}}  \prod_{i=1}^n \hat{A}(\xi_{\sigma(i)}, \xi_{\sigma(i+1)}),
\end{split}
\ee
where $\hat{A}(\xi_i, \xi_j)$ are defined by:
\be
\hat{A}(\xi_i, \xi_j) = \begin{cases}
i_{\xi_i, \xi_j} \frac{1}{\xi_i-\xi_j} + A(\xi_i, \xi_j),  & i < j, \\
A(\xi_i, \xi_i),  & i =j, \\
i_{\xi_j, \xi_i} \frac{1}{\xi_i-\xi_j} + A(\xi_i, \xi_j),  & i > j.
\end{cases}
\ee
Here the following convention is used: $\sigma(n+1) = \sigma(1)$.
\end{thm}

\section{DJKM Construction of Lie Subalgebra of the Lie Algebra $A_\infty$}

\label{sec:DJKM}

Starting from this section we will list some examples for which the method of last section can be applied.

In this section we recall the construction of some Lie subalgebras of $A_\infty$
by the Japanese school \cite{DJKM} and \cite{Jimbo-Miwa}  based on automorphisms of $A_\infty$ and their fixed point sets.
We will follow closely the notations in these two references.
Their notations are different from our notations used in last section.
The translation from their notations to ours is given as follows:
\bea
&& \psi_i, \;\;\; i \in \bZ \rightleftharpoons \psi_r= \psi_{-i - 1/2}, \;\;\; r \in \half + \bZ, \\
&& \psi_i^*, \;\;\; i \in \bZ \rightleftharpoons \psi^*_s= \psi^*_{i + 1/2}, \;\;\; s \in \half + \bZ.
\eea
Denote by $\bA$ the Clifford algebra generated by $\{ \psi_i, \psi_i^*\}_{i \in \bZ}$.

\subsection{The Lie algebra $A_\infty$}

The Lie algebra $A_\infty$ or $\gl(\infty)$ is defined by
\be
\begin{split}
A_\infty =\{ X = & \sum_{i, j \in \bZ} a_{ij} : \psi_i \psi_j^*: + \lambda \:|\;
\text{there exists an $N$} \\
& \text{such that $a_{ij} = 0$ for $|i-j|> N$ } \}.
\end{split}
\ee
Let $H_0 = \sum_{i\in \bZ} :\psi_i\psi_i^*:$.
For $l \in \bZ$,
let $\cF^{(l)} =\{ v\in \cF\;|\; H_0v = l \cdot v\}$.
Then $A_\infty$ acts on each $\cF^{(l)}$,
giving rise to an irreducible representation of $A_\infty$.
A Chevalley basis is given by
\be
e_i = \psi_{i-1} \psi_i^*, \;\;
f_i = \psi_i \psi_{i-1}^*, \;\;
h_i = \psi_{i-1} \psi_{i-1}^* - \psi_i\psi_i^*, \;\; \psi_0\psi_0^*.
\ee
The vector $|l \rangle$ defined by
\be
|l\rangle =
\begin{cases}
\psi_l^* \cdots \psi_{-1}^*\vac & (l < 0), \\
\vac & (l=0), \\
\psi_{l-1} \cdots \psi_0 \vac & (l > 0)
\end{cases}
\ee
gives the highest weight vector of $\cF^{(l)}$:
\be
e_i |l\rangle = 0, \;\; h_i|l\rangle = \delta_{il} |l\rangle, \;\; i \in \bZ.
\ee

boson-fermion correspondence

\subsection{The Lie algebra $B_\infty$ and $C_\infty$}
Consider the automorphisms $\sigma_l$ of $A_\infty$ induced by
\be
\begin{split}
\sigma_l(\psi_n) = (-1)^{l-n} \psi^*_{l-n}, \\
\sigma_l(\psi^*_n) = (-1)^{l-n} \psi_{l-n}.
\end{split}
\ee
The subalgebras $B_\infty$ and $C_\infty$ in $A_\infty$ are defined
as the fixed point set of $\sigma_0$ and $\sigma_1$ respectively:
\bea
&& B_\infty = \{ X \in A_\infty \;|\; \sigma_0(X) = X\}, \\
&& C_\infty = \{ X \in A_\infty \;|\; \sigma_1(X) = X\}.
\eea
When restricted to $B\infty$,
each $\cF^{(l)}$ is an irreducible highest weight $B_\infty$-module with highest weight vectors $|l \rangle$,
whose weight is given:
\be
wt(|l\rangle)
= \begin{cases}
\Lambda_{l-1}, & l \geq 2, \\
2 \Lambda_0, & l =0 ,1, \\
\Lambda_{-l}, & l \leq -1.
\end{cases}
\ee
On the other hand,
as a $C_\infty$-module $\cF^{(l)}$ is no longer irreducible.
Nevertheless, $|l\rangle$ generates a highest weight module $V_l$ with weight
\be
wt(|\l\rangle) = \begin{cases}
\Lambda_l, & l \geq 0, \\
\Lambda_{-l}, & l < 0,
\end{cases}
\ee
and $\cF^{(l)}$ splits as follows:
\be
\cF^{(l)} \cong \cF^{(-l)} \cong V_l \oplus V_{l+2} \oplus V_{l+4} \oplus\cdots.
\ee

\subsection{The Lie algebra $B_\infty'$}

In last subsection,
we have seen that the restriction of $\cF^{(0)}$ to $B_\infty$ has highest weight $2 \Lambda_0$.
There is another realization of $B_\infty$  that leads to a highest weight representation
with highest weight $\Lambda_0$.

Define two sets of neutral fermions as follows:
\be
\phi_m = \frac{\psi_m + (-1)^m \psi_{-m}^*}{\sqrt{2}}, \;\;\;
\hat{\phi}_m = i \frac{\psi_m - (-1)^m \psi_{-m}^*}{\sqrt{2}}, \;\;\; (m \in \bZ).
\ee
They satisfies the following anti-commutation relations:
\begin{align*}
[\phi_m, \phi_n]_+ & = (-1)^m \delta_{m, -n}, &
[\hat{\phi}_m, \hat{\phi}_n]_+ & = (-1)^m \delta_{m, -n}, \\
[\phi_m, \hat{\phi}_n]_+ & = 0, & m, n \in \bZ.
\end{align*}
Define an automorphism $\kappa: \bA \to \bA$ by
\begin{align}
\kappa(\phi_m) & = \hat{\phi}_m, & \kappa(\hat{\phi}_m) = - \phi_m.
\end{align}
The Lie algebra
$$B_\infty' = \{ \sum a_{ij} : \phi_i\phi_j: \;\; |\;\;
\text{there exists $N$, $a_j  = 0$ if $|i+j|> N$}\}$$
is isomorphic to $B_\infty$ by the following map
\be
B_\infty' \to B_\infty, \;\;\; X \mapsto X + \kappa(X).
\ee
The vector $\vac$ generates a highest weight $B_\infty'$-module
with highest weight $\Lambda_0$.

The Lie algebra $B'_\infty$ does not belong to $A_\infty$,
so we cannot directly apply the method of \S \ref{sec:KP}.
Nevertheless,
the tau-function for $B_\infty$ and $B_\infty'$ are related as follows:
\be
\tau_{B_\infty'}(T_1, T_3, \dots)^2
= \tau_{B_\infty}(T_1, T_2, T_3, T_4, \dots)|_{T_2 =T_2 = \cdots =0},
\ee
and for the latter we can apply the method of \S \ref{sec:KP}.

\subsection{The Lie algebra $D_\infty'$}

The Lie algebra $D'_\infty$ is defined as follows
\be
\begin{split}
D_\infty' = & \{ \sum a_{jk} :\psi_j\psi_k^*: + b_{jk} \psi_j\psi_k
+ c_{jk} \psi_j^*\psi_k^* + d \\
& \text{$\exists N$, $a_{jk} = b_{j,k}= c_{j, k} = 0$ if $|j+k| > N$} \}.
\end{split}
\ee
As a D$\infty$-module,
$\cF$ splits into two irreducible highest weight modules
generated by $\vac$ and $|1\rangle$, respectively,
and their highest weights are $\Lambda_0$
and $\Lambda_1$, respectively.

\subsection{The Lie algebra $D_\infty$}

The two-component charged free fermions are defined by:
\begin{align}
\psi_n^{(1)} & = \psi_{2n}, & \psi_n^{(2)} & = \psi_{2n+1}, \\
\psi_n^{(1)*} & = \psi^*_{2n}, & \psi_n^{(2)*} & = \psi^*_{2n+1}.
\end{align}
Denote by $\sigma$ the automorphism of the Clifford algebra of the two-component
charged free fermions  given by
\be
\sigma(\psi^{(j)}_n) = (-1)^n \psi^{(j)*}_{-n}, \;\;\;\;
\sigma(\psi_n^{(j)*}) = (-1)^n \psi^{(j)}_{-n}, \;\;\; j =1,2.
\ee
Then we define
\be
D_\infty = \{ X \in A_\infty \;|\; \sigma(X) = X\}.
\ee
The fermionic Fock space splits into highest weight representations with highest weights
$\Lambda_0+\Lambda_1$, $2\Lambda_0$, $2\Lambda_1$, $\Lambda_j$ ($j \geq 2$).

\subsection{Reduction to Kac-Moody algebras }

We call $X = \sum_{i,j\in \bZ} a_{ij} :\psi_i\psi_j^*: +c \in A_\infty$ $l$-reduced
if and only if the following
conditions (i) and (ii) are satisfied:
\begin{itemize}
\item[(i) ] $a_{i+l,j+l} = a_{i,j}$, $i, j \in \bZ$,
\item[(ii)] $\sum_{i = 0}^{l-1} a_{i,i+jl} = 0$,  $(j\in \bZ$).
\end{itemize}
We call $X=  \sum_{\mu, \nu = 1,2} \sum_{i, j \in \bZ} a_{i,j}^{(\mu, \nu)} :\psi_i^{(\mu)}\psi_j^{(\nu)*}:
+c \in A_\infty$ \; $(l_1, l_2)$-reduced if and only if the following conditions
(i)$'$ and (ii)$'$ are satisfied.
\begin{itemize}
\item[(i)$'$] $a^{(\mu, \nu)}_{i+l_\mu, j + l_\nu} = a^{(\mu, \nu)}_{i, j}$, $\mu, \nu = l,2$, $i,i \in \bZ$),
\item[(ii)$'$] $\sum_{\mu=1,2} \sum_{i=0}^{l_\mu-1} a_{i, i+jl_\mu}^{(\mu, \nu)} = 0$,
 ($j \in \bZ$).
\end{itemize}

The Lie subalgebras $A^{(1)}_l$, $D^{(2)}_l$, $A_{2l}^{(2)}$, $C^{(1)}_l$,  $D^{(1)}_l$ and $A^{(2)}_{2l-1}$
are obtained as follows:
\ben
A_l^{(1)} & = & \{ X \in A_\infty \;|\; X: (l + 1)-\text{reduced} \}, \\
D_{l+1}^{(2)} & = & \{ X \in B_\infty \;|\; X: 2(l + 1)-\text{reduced} \} = A_{2l+1}^{(1)} \cap B_\infty \\
& \cong & \{ X \in C_\infty \;|\; X: 2(l + 1)-\text{reduced} \} = A_{2l+1}^{(1)} \cap C_\infty, \\
C_{l}^{(1)} & = & \{ X \in C_\infty \;|\; X: 2l-\text{reduced} \} = A_{2l}^{(1)} \cap B_\infty, \\
D_{l}^{(1)} & = & \{ X \in D_\infty \;|\; X: (2l-2s, 2s)-\text{reduced} \}, \;\; 1 \leq s \leq l-1, \\
A_{2l-1}^{(2)} & = & \{ X \in D_\infty \;|\; X: (2l-2s-1, 2s+1)-\text{reduced} \},  \\
&&  \;\; 0 \leq s \leq l-1.
\een
According to \cite{DJKM},
the reduction can be used to obtain the principal realization of the basic representations of these Lie algebras
given in \cite{KKLW}.

\subsection{Other cases of affine Kac-Moody algebras}

We conjecture the principal realization of basic representation for the basic representations of
other Lie algebras given in \cite{KKLW} can be obtained by a reduction of $A_\infty$,
in particular,
$E_n^{(1)}$ ($n=6,7,8$).
If this is true,
then from the constructed embedding $X^{(1)}_n \subset A_\infty$ ($X=A, D, E$),
one can construct embeddings $Y^{(k)}_n \subset A_\infty$ for all affine Kac-Moody algebras
by the following well-known constructions using automorphisms of the extended Dynkin diagrams of $X^{(1)}_n$.
Suppose that  the extended Dynkin diagram of $\fg$ admits an automorphism $\sigma$ of order $k > 1$
that preserves the vertex $\alpha_0$.
Then $\sigma$ induces an automorphism of $\fg$ of order $k$.
Let
\be
\fg = \oplus_{j=0}^{k-1} \fg_j,
\ee
where $\sigma$ has eigenvalue $e^{2\pi ij/k}$ on $\fg_j$.
Then $\fg_0$ is also a simple Lie algebra.
The following are all the possible cases:
Case 1. $\fg = D_{n+1}$,  $k=  2$, $\fg_0 = B_n$;
Case 2.  $\fg=  A_{2n-1}$, $k= 2$,$\fg_0 = C_n$;
Case 3. $\fg = E_6$, $k=2$, $\fg_0 = F_4$;
Case 4.  $\fg =D_4$, $k=3$,  $\fg_0 = G_2$.
Let $\hat{\fg}^{(1)}$ be the affine Kac-Moody algebra associated to $\fg$.
Consider the automorphism $\hat{\sigma}$ of $\fg^{(1)}$ induced by $\sigma$ defined by
\begin{align*}
e_i & \mapsto e_{\sigma(i)}, & f_i & \mapsto f_{\sigma(i)}, &
h_i & \mapsto h_{\sigma(i)}.
\end{align*}
Then the fixed point set of $\hat{\sigma}$ is the nontwisted affine Kac-Moody algebra $(\fg_0)^{(1)}$
associated to $\fg_0$.
One can define another automorphism $\tilde{\sigma}$ on $\fg^{(1)}= \fg[z,z^{-1}] \oplus \bC c \oplus \bC d$ by
\begin{align*}
\tilde{\sigma}(z^n \otimes X) &= e^{-n \cdot 2\pi i /k} \cdot \sigma(X), \; X \in \fg, &
\tilde{\sigma}(c) & = c, & \tilde{\sigma}(d) & = d.
\end{align*}
Then the fixed point set of $\tilde{\sigma}$ is the twisted affine Kac-Moody algebra $\fg^{(k)}$.
Since $\fg^{(1)}$ is embedded in $A_\infty$,
so are $(\fg_0)^{(1)}$ and  $\fg^{(k)}$
as they are subsets of $\fg^{(1)}$.

\section{Matrix Construction of Lie Subalgebra of the Lie Algebra $A_\infty$}

\label{sec:Matrix}

In this section we explain that  affinization of embedding into $gl_n$ can be used construct
many subalgebras of $A_\infty$.
Depending on the embedding index,
the representation induced from $\cF$ may have level $>1$.

\subsection{Matrix realizations and reductions}

We first realize the Lie algebra $A_{\infty}$ in terms of infinite matrices.
\be
\bar{A}_\infty = \{ (a_{ij})_{i,j\in \bZ} \;|\; a_{ij} = 0 \;\; \text{if $|i-j| \gg 0$} \}.
\ee
Denote by $E_{ij}$ the matrix with $1$ as the $(i,j)$ entry and all other entries $0$.
Because one clearly has:
\be
E_{ij} E_{kl} = \delta_{jk} E_{il},
\ee
therefore one gets:
\be
[E_{ij}, E_{kl}] = \delta_{jk} E_{il} - \delta_{li} E_{kj}.
\ee
The Lie algebra $A_\infty$ as a vector space is
\be
A_\infty = \bar{A}_\infty \oplus \bC c,
\ee
but with the following commutation relations:
\be
[E_{ij}, E_{kl}] = \delta_{jk} E_{il} - \delta_{li} E_{kj} + \alpha(E_{ij}, E_{kl}) \cdot c,
\ee
where $\alpha$ is a cocycle defined by:
\be\begin{split}
& \alpha(E_{ij}, E_{ji}) = - \alpha(E_{ji}, E_{ij}) = 1, \;\; \text{if $i \leq 0$, $j \geq 1$}, \\
& \alpha(E_{ij}, E_{kl}) = 0, \;\;\; \text{in all other cases}.
\end{split}
\ee
There is a natural representation of $A_\infty$ on $\cF$ defined as follows:
\be
\begin{split}
& \hat{r}(E_{ij}) = :\psi_i\psi_j^*:, \\
& \hat{r}(c) = 1.
\end{split}
\ee

\subsection{An embedding of $\widehat{\gl'_n}$ in $A_\infty$}

Let $\gl_n[t,t^{-1}]$ be the set of Laurent polynomials with coefficients in $\gl_n$.
An element of $\gl_n[t,t^{-1}]$  has the form
\be
a(t) = \sum_{k\in\bZ} t^ka_k (a_k \in \gl_n) ,
\ee
where $a_k=0$ for $k\gg 0$ or $k \ll 0$.
Denote by $e_{i,j}$ the $n\times n$ matrix which has $1$ as the $(i,j)$ entry and $0$ elsewhere.
It is clear that the matrices
\be
e_{ij}(k) := t^k e_{ij}, \;\;
(1 \leq i,j \leq n,\; k \in \bZ)
\ee
form a basis of $\gl_n$.

The Lie algebra $\gl_n$ acts on the vector space  $\bC^n$
which has a standard basis $u_1, \dots, u_n$.
This induces an action of $\gl_n[t,t^{-1}]$ on $\bC[t, t^{-1}]^n$,
The vectors
\be
v_{nk+j} := t^{-k}  u_j
\ee
form a basis of $\bC[t, t^{-1}]^n$.
It is clear that
\be
e_{ij}(k) v_{nl+j} = v_{n(k-l)+i}.
\ee
Therefore, the action of $e_{ij}(k)$ on $\bC[t,t^{-1}]^n$ after the identification
with $\bC^\infty$ can be represent by a matrix in $\bar{A}_\infty$:
\be
R(e_{ij}(k)) = \sum_{l \in \bZ} E_{n(l-k)+i, nl+j}.
\ee
For $a(t) \in \gl_n[t,t^{-1}]$,
denote the corresponding matrix in $\bar{A}_\infty$ by  $R(a(t))$.
It has the following block form:
\be
R(a(t))
= \begin{pmatrix}
\dots &  \dots & \dots & \dots & \dots & \dots & \dots \\
\dots & a_{-1} & a_0   & a_1   & \dots & \dots & \dots \\
\dots & \dots & a_{-1} & a_0   & a_1   & \dots & \dots \\
\dots & \dots & \dots & a_{-1} & a_0   & a_1   & \dots \\
\dots &  \dots & \dots & \dots & \dots & \dots & \dots
\end{pmatrix}
\ee
It can be checked that $R: \gl_n[t,t^{-1}] \to \bar{A}_\infty$
is an injective Lie algebra homomorphism.
One finds
\be
\alpha(R(e_{ij}(k)), R(e_{pq}) = \delta_{iq} \delta_{jp} \delta_{l+k,0}.
\ee
It follows that for  general elements $a(t), b(t) \in gl_n[t, t^{-1}]$,
\be
\alpha(a(t), b(t)) =  \res ( a'(t) b(t) ).
\ee

The Lie algebra $\widehat{\gl'_n}$ is the vector space
\be
\widehat{\gl'_n} = \gl_n[t,t^{-1}] \oplus \bC c,
\ee
with the following commutation relations:
\be
\begin{split}
& [c, a(t)] = 0, \\
& [a(t), b(t)] = a(t)b(t) - b(t)a(t) + \res (a'(t)b(t)) \cdot c.
\end{split}
\ee
One can extend $R$ to an injective homomorphism $\hat{R}: \widehat{\gl'_n} \to A_\infty$
as follows:
\be
\hat{R}(a(t) + \lambda c) = R(a(t)) + \lambda c.
\ee

\subsection{Affinization of embeddings}

Suppose that $\fg$ is simple Lie algebra and $\iota: \fg \hookrightarrow \gl_n$
is an embedding of Lie algebra.
Then it induces an embedding of
$\tilde{\iota}: \fg[t, t^{-1}] \to \gl_n[t, t^{-1}]$.
Define $\alpha_\iota: \fg[t,t^{-1}] \otimes \fg[t, t^{-1}] \to \bC$ by
\be
\alpha_\iota( X(t), Y(t)) = \alpha (R(\tilde{\iota}(X(t))), R(\tilde{\iota}(Y(t))) ),
\ee
and define a central extension $\hat{\fg}_\iota$ of $\fg[t, t^{-1}]$,
by $\alpha_\iota$:
\be
[X(t)+\lambda c, Y(t) + \mu c]
= [X(t), Y(t)] + \alpha_\iota(X(t), Y(t)) \cdot c.
\ee
One can also consider the twisted construction when there is a nontrivial automorphism
of the extended Dynkin diagram preserving the vertex $\alpha_0$.
This yields an embedding of $\hat{\fg}^{(k)}$ in $\widehat{\gl'_n}$.

\section{Concluding Remarks}

We have shown that there are many examples of Lie subalgebras of $A_\infty$
that lead to integrable hierarchies that one can apply
our general results for KP hierarchy.
Most interesting cases are those arising from FJRW theory \cite{FJR, LRZ},
whose partition functions are tau-functions of the Drinfeld-Sokolov hierarchy 
of type A-G and satisfy the puncture equation (see e.g. \cite{Li, Cafasso-Wu}):
\be
\biggl( \sum_{i\in E_+} \biggl(\frac{i + h}{h} t_{i+h} - \delta_{i,1}\biggr) \frac{\pd}{\pd t_i}
+ \frac{1}{2h} \sum_{i,j \in E_+^0; i+j =h}
ijt_i t_j \biggr)  \tau(t) = 0,
\ee
where $h =\sum_{i=0}^k k_i$  is the Coxeter number for
the untwisted affine Kac-Moody algebra $\hat{\fg}$.
We have seen that for most of them their tau-functions and $n$-point functions 
can be found in a uniform way in a fermionic picture by treating them as reductions 
of the KP hierarchy and apply the general method developed in \cite{Zhou-Emergent}.
We conjecture this method actually applies to all of them.
The missing cases are $E_6^{(1)}$, $E_7^{(1)}$, $E_8^{(1)}$ and $F_4^{(1)}$ at present.

\end{document}